\documentclass[aps,prl,twocolumn,twoside,floatfix]{revtex4}
\usepackage{graphicx}
\usepackage{color}

\begin{document}

\newcommand{\Al}{Al$^+$ }
\newcommand{\Alns}{Al$^+$}
\newcommand{\AlTS}{$^{27}$Al$^+$ }
\newcommand{\AlTSns}{$^{27}$Al$^+$}
\newcommand{\Be}{Be$^+$ }
\newcommand{\Bens}{Be$^+$}
\newcommand{\BeN}{$^{9}$Be$^+$ }
\newcommand{\BeNns}{$^{9}$Be$^+$}
\newcommand{\Mg}{Mg$^+$ }
\newcommand{\Mgns}{Mg$^+$}
\newcommand{\MgTF}{$^{25}$Mg$^+$ }
\newcommand{\MgTFns}{$^{25}$Mg$^+$}
\newcommand{\Hg}{Hg$^+$ }
\newcommand{\clocktransition}{$^1$S$_0$$\leftrightarrow$$^3$P$_0$ }
\newcommand{\clocktransitionAlmF}[2]{$|^1S_0, m_F=#1\rangle$$\leftrightarrow$$|^3P_0, m_F=#2\rangle$}
\newcommand{\diffAlAl}{$-1.8\times10^{-17}$ }
\newcommand{\diffpmAlAl}{$(-1.8\pm0.7)\times10^{-17}$ }
\newcommand{\diffpmAlAlns}{$(-1.8\pm0.7)\times10^{-17}$}
\newcommand{\diffAlAlns}{$-1.8\times10^{-17}$}
\newcommand{\sigmaAlAl}{$2.5\times10^{-17}$ }
\newcommand{\uncStat}{$7.0\times10^{-18}$ }
\newcommand{\uncStatns}{$7.0\times10^{-18}$}
\newcommand{\systAlMg}{$8.6\times10^{-18}$ }
\newcommand{\systAlMgns}{$8.6\times10^{-18}$}
\newcommand{\stabAlMg}{$2.8\times10^{-15}$}
\newcommand{\systAlBe}{$2.3\times10^{-17}$ }
\newcommand{\systSr}{$1.5\times10^{-16}$ }
\newcommand{\systHg}{$1.9\times10^{-17}$ }
\newcommand{\dnn}{\Delta\nu/\nu}
\newcommand{\xfertransition}{$^1$S$_0$ $\rightarrow$ $^3$P$_1$ }
\newcommand{\MMns}{EMM}
\newcommand{\MM}{EMM }
\date{\today}



\title{Frequency Comparison of Two High-Accuracy Al$^+$ Optical Clocks}


\author{C. W. Chou}
\email[]{chinwen@nist.gov}
\author{D. B. Hume}
\author{J. C. J. Koelemeij}
\thanks{Present address: Laser Centre Vrije Universiteit Amsterdam, The Netherlands}
\author{D. J. Wineland}
\author{T. Rosenband}


\affiliation{Time and Frequency Division, National Institute of Standards and Technology, Boulder, Colorado 80305}

\begin{abstract}
We have constructed an optical clock with a fractional frequency inaccuracy of $8.6\times10^{-18}$, based on quantum logic spectroscopy of an Al$^+$ ion.  A
simultaneously trapped Mg$^+$ ion serves to sympathetically laser-cool the
Al$^+$ ion and detect its quantum state. The frequency of the
$^1$S$_0$$\leftrightarrow$$^3$P$_0$ clock transition is compared to that of a previously constructed Al$^+$ optical clock with a statistical measurement uncertainty of $7.0\times10^{-18}$. The two clocks exhibit a relative stability of $2.8\times10^{-15}\tau^{-1/2}$, and a fractional frequency difference of $-1.8\times10^{-17}$, consistent with the accuracy limit of the older clock.
\end{abstract}

\maketitle

Optical clocks based on petahertz ($10^{15}$ Hz) transitions in
isolated atoms have demonstrated  significant improvements over the current cesium
primary-frequency-standards at 9.2 GHz.  They also shed light on fundamental
physics, such as the possible variation of physical constants.
While the merits of laser-cooled ion optical frequency standards
were known~\cite{HGD1982monoion,DJW1987LaserCoolingLimits},
further developments were required to permit their use.
Sub-hertz linewidth lasers~\cite{BCY1999subhertz}
enabled single ions to be probed with sufficient resolution for high-stability clock operation, and
control of external-field perturbations allowed such clocks to operate with an inaccuracy below $10^{-16}$~\cite{HgCs2006,TR2008AlHg}.
For comparison, cesium standards that realize the SI second have reached an inaccuracy of $3\times10^{-16}$~\cite{2008ParkerEFTF},
and an optical lattice clock based on Sr atoms has been reported~\cite{Campbell2008} with an
inaccuracy of $1.5\times10^{-16}$.  Here we describe an \Al ion clock with an inaccuracy of \systAlMgns.

The \clocktransition transition in \Al at 1.121 PHz is of interest
due to its low sensitivity to electromagnetic perturbations and
its narrow natural linewidth of 8 mHz
\cite{Dehmelt1992,TR2007Al3P0observed}. \Al has the
smallest sensitivity to blackbody radiation
\cite{TR2006BBRshift,Mitroy2009BBR} among atomic species currently under
consideration for clocks, thus relaxing the requirement on ambient
temperature control.  However, the
absence of an accessible allowed optical transition prevents the
internal state of \Al ions from being detected by conventional methods,
and the ion cannot currently be directly laser-cooled. Quantum logic spectroscopy
(QLS)~\cite{POS2005BeAl} overcomes these difficulties by trapping a
``logic ion'' that can be directly laser-cooled together with the \Al
clock ion.  The coupled motion of the two ions allows for sympathetic
cooling, as well as the transfer of the clock ion's quantum state to the
logic ion, where the state can be measured.

The clock described here shares features with our previously-constructed Al-Be
clock~\cite{TR2007Al3P0observed,TR2008AlHg},
but also includes many changes, making a
comparison of the two clocks a valuable test of systematic errors.
  In the new clock, the \BeN logic ion has been replaced by \MgTFns,
whose mass closely matches that of \AlTSns.  Laser-cooling
inefficiencies due to mass-mismatch are thus suppressed.  The Al-Mg ion
trap is a linear Paul trap built from tool-machined all-metal electrodes
(Fig. 1).  This construction differs from the Al-Be trap that was
built from laser-machined and gold-coated alumina electrodes
\cite{Rowe2002}, and it exhibits reduced RF-micromotion-inducing electric
fields.

\begin{figure}
\includegraphics[angle=0, width=0.5\textwidth]{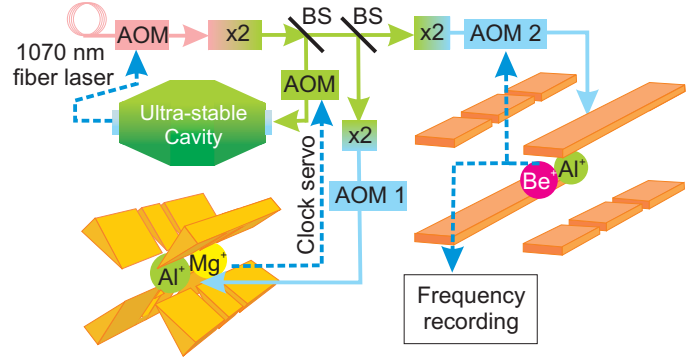}
\caption[Experimental setup.]{Setup for comparing the frequencies of the
two Al$^+$ clocks. The 4$^{th}$ harmonic of a fiber laser is locked to the
\Al \clocktransition
clock transition in the Al-Mg apparatus with a fixed offset frequency (applied to AOM1). Another laser beam derived from the same laser probes the clock transition in the Al-Be apparatus, where the laser frequency is locked to the clock transition in a separate digital feedback loop that controls AOM2. The record of frequencies applied to AOM2 represents the difference in clock frequencies. BS: beam splitter; AOM: acousto-optic
modulator; x2: frequency doubler.}
\label{Setup}
\end{figure}

QLS with a \Mg logic ion proceeds in the same way as with \Bens,
but the ground-state-cooling process
\cite{CM1995GScooling} has been modified to require only two lasers
rather than three~\cite{TwoLaserGSC}.  This
cools the out-of-phase axial motional mode to an
average quantum number of $\bar{n} < 0.05$.  It also enables
quantum-non-demolition transfer (QNDT) of the \Alns-clock-ion state to \Mg
with approximately 80~\% fidelity in a single QNDT repetition and over
99~\% fidelity after typically five QNDT repetitions
\cite{DBH2007detection}.

The trap utilized in the new \Al clock has blade-shaped
gold-coated beryllium-copper electrodes (Fig.~\ref{Setup}) whose
edges are approximately 400~$\mu$m from the ions. The \MgTF logic ion ($I=5/2$) is
manipulated with light of 279.5~nm wavelength from two independent lasers.  A
frequency-doubled dye laser resonantly drives
$|^2S_{1/2},F\in\{2,3\}\rangle$ $\rightarrow$ $^2P_{3/2}$ cycling and
repumping transitions, while a frequency-quadrupled fiber laser is
detuned by 40-60 GHz to drive Raman transitions between the \MgTF
hyperfine qubit states.  Both ions are created via multi-photon-ionization of neutral atoms from
ovens.  A 396 nm diode laser produces \AlTSns~\cite{Hurst1979PI}, while a
frequency-doubled dye laser at 285 nm creates \MgTF~\cite{Drewsen2000PI}. Typically the clock transition is probed with 150~ms duration $\pi$-pulses with a duty cycle of approximately 65~\%.  The remaining 35~\% is occupied with state preparation and state detection functions as well as interleaved experiments that allow real-time measurement and reduction of micromotion.

\begin{table}
\caption{Systematic effects that shift the clock from its
ideal unperturbed frequency. Shifts and uncertainties given are in fractional frequency units ($\dnn$).  See text for discussion.}
\begin{tabular}{l r r}
\hline
Effect & Shift & Uncertainty\\
        & ($10^{-18}$) & ($10^{-18}$) \\
\hline
Excess micromotion &  -9 & 6\\
Secular motion & -16.3 & 5\\
Blackbody radiation shift &  -9 & 3 \\
Cooling laser Stark shift & -3.6 & 1.5\\
Quad. Zeeman shift & -1079.9 & 0.7\\
Linear Doppler shift & 0 & 0.3\\
Clock laser Stark shift & 0 & 0.2\\
Background gas collisions & 0 & 0.5\\
AOM freq. error & 0 & 0.2\\
\hline
Total & -1117.8 & 8.6\\
\end{tabular}
\label{systematics}
\end{table}

Systematic shifts of the Al-Mg clock are listed in Table~\ref{systematics}.
Two types of residual motion cause time-dilation shifts: micromotion
near the trap drive frequency of $\nu_{RF}=59$ MHz, and harmonic-oscillator
(secular) motion at the ion's normal mode frequencies (Table~\ref{ModesTable}).  In both cases the clock frequency shifts by
$\frac{\Delta\nu}{\nu}=-\frac{\langle v^2\rangle}{2c^2}\left(1+(\frac{f}{\mathrm{400 MHz}})^2\right)$, where we add to the relativistic time-dilation $\langle v^2\rangle/(2c^2)$ a frequency-dependent term that corresponds to the Stark shift from the motion-inducing electric fields.  Here $\langle v^2 \rangle$ is the mean-squared \Alns-ion velocity and $f$ is the frequency of motion.  For the highest motional frequencies the Stark shift correction is 2~\%.

\begin{table}\caption{Motional modes of the Al-Mg ion pair.  For each of
the six normal modes, the oscillation frequency and zero-point motional
amplitude for each ion is listed.  The trap axis corresponds to
$\hat{z}$, and $\hat{x}, \hat{y}$ are two orthogonal radial directions
whose orientation is determined by the trap geometry. Also shown are the
calculated ($\bar{n}_C$) and measured ($\bar{n}_M$) Doppler-cooled
average motional quantum numbers, and the time-dilation (TD) per
motional quantum as well as the total TD per mode.}
\begin{tabular}{rc|cccccc}
\hline
f & [MHz] &  6.53 &  5.66 &  5.20 &  4.64 &  3.41 &  3.00 \\
\hline
\MgTF & [nm] & 4.9$\hat{y}$ & 2.9$\hat{y}$ & 4.6$\hat{z}$ &5.5$\hat{x}$ &-4.2$\hat{x}$ & 5.6$\hat{z}$ \\
\AlTS & [nm] & 2.6$\hat{y}$ & -5.0$\hat{y}$ & -4.1$\hat{z}$ &3.5$\hat{x}$ & 6.2$\hat{x}$ &   5.8$\hat{z}$\\
\hline
$\bar{n}_M$ &            & 2.9 &  4.5 &  3.4 &  6.3 &  10.0 &  7.0 \\
$\bar{n}_C$ &            & 3.3 &  3.8 &  3.4 &  5.9 &  8.0 &  5.9 \\
TD/quantum & $[10^{-18}]$ & 0.226 & 0.731 & 0.197 & 0.290 & 0.771 & 0.133 \\
Total TD  & $[10^{-18}]$ &  0.77 &  3.66 & 0.77 &  1.97 & 8.10  & 1.00 \\
\hline
\end{tabular}
\label{ModesTable}
\end{table}

Excess micromotion (\MMns) refers to the rapid ion motion at $\nu_{RF}$~\cite{Berkeland1998MM}.
It is caused by electric fields that force the ion away from the RF-minimum of
the ion trap, or phase shifts between trap electrodes that cause the RF-fields to be non-zero
at the pseudo-potential minimum.  We measure the amplitude of this motion at $\nu_{RF}$ by observing
the motional-sideband strength of the \Al \xfertransition transition in
three orthogonal directions.  For small amplitude of motion, the time
dilation shift is $\dnn = -|\vec{\eta} \nu_{RF}/\nu_L|^2 =
-2.8\times10^{-15}|\vec{\eta}|^2$, where $\nu_L = 1.12$ PHz is the probe laser
frequency, and $\vec{\eta}=(\eta_1,\eta_2,\eta_3)$ is the measured \MM
Lamb-Dicke parameter (the ratio of sideband and carrier Rabi rates) for
the three orthogonal directions.  Typical values for $\eta_{1,2,3}$ are
0.01 to 0.04.  We note that this \MM measurement method detects slow
electric field fluctuations such as those caused by the migration of
photo-electrons, as well as faster fluctuations caused by periodic line
noise (50 to 60 Hz).  Fluctuations that are shorter than the laser probe
period of 0.05 to 0.1 ms will not be detected.

In a perfect linear Paul trap \MM along the trap axis does not occur, but imperfections in the
ion-trap geometry can lead to axial \MM.  For the Al-Mg trap there exists
a sharp minimum of axial \MM at one spatial location.  The \Al ion is
held at this point, but random background-gas collisions cause the Al-Mg
ion pair to spontaneously re-order, which causes the \Al ion to move by 3~$\mu$m every 200~s, on
average.  The \Al ion is maintained at the position of minimal
micromotion by adjusting the electrode voltages every 10 s to force the
ion-pair into the desired order. When the ions are in the wrong order
(about 5~\% of the time), the \Al ion experiences excess axial micromotion
and a corresponding clock shift of $\dnn=-2.7\times10^{-17}$.  This additional shift is included in
Table~\ref{systematics}.

During each \Al clock interrogation pulse, the \Mg ion is simultaneously
Doppler cooled by a laser that is tuned 21 MHz below the
$|^2S_{1/2},F=3,m=-3\rangle$ $\rightarrow$ $|^2P_{3/2},F=4,m=-4\rangle$
cycling transition.  The amplitude of secular motion (corresponding to the motional
temperature) is extracted from the ratio of amplitudes
for the red- and blue-sidebands of a Raman transition for all six normal
modes~\cite{CM1995GScooling}.  Measured values are shown in
Table~\ref{ModesTable}, together with values calculated from
laser-cooling theory for a single ion~\cite{Itano1982Cooling}.
The single-\Mgns-ion Doppler cooling limit is also valid when applied to each of the six normal modes of
the Al-Mg ion pair.  The measured and calculated motional quantum numbers agree within the measurement uncertainty,
and we consider the stated 30~\% uncertainty for this shift to be a conservative limit.

The \Mg Doppler cooling laser beam maintains the Al-Mg ion pair at a constant
motional temperature during the clock interrogation pulse, but because it also overlaps the \Al ion, it
causes an AC Stark shift by coupling off-resonantly to allowed
transitions that connect to the ground (3s$^2)$$^1$S$_0$ and excited (3s3p)$^3$P$_0$ clock states.  Following
the evaluation of the blackbody radiation shift~\cite{TR2006BBRshift},
we estimate the differential clock polarizability at $279.5$~nm as
$\dnn=(-3.5\pm0.6)\times10^{-17}S$, where ($S=I/I_S$) is the saturation
parameter for \MgTF ($I_S \approx 2470$~W/m$^2$).  The intensity $I$ of the \Mg
Doppler cooling laser is estimated from the rate at which this laser
repumps the $|^2S_{1/2},F=2,m=-2\rangle$ dark hyperfine ground state.
The ion fluorescence photo-multiplier counts $F(t)$ collected in a duration $t$
due to repumping of the dark
state may be written as $F(t)=b(t+\tau(e^{-t/\tau}-1))$ where
$\tau=(0.217/S)$ ms and $b$ is the bright-state counting rate.  We extract $S$ by fitting $F(t)$
to the observed ion fluorescence.  Typically, we measure $\tau=2.1\pm0.8$ ms,
and find $\dnn=(-3.6\pm1.5)\times10^{-18}$.

Another Stark shift is caused by thermal blackbody radiation.
The temperature of the Al-Mg ion trap is measured with two platinum
sensors at opposite ends of the trap structure which are expected to be
at temperature extremes.  Heat is removed
primarily through thermal conduction at one end of the trap
where we measure a temperature of 35~$^\circ$C.  At the higher thermal resistance
trap-end the temperature is 40~$^\circ$C, and the thermal radiation field
impinging upon the ion is bounded by this maximum temperature and the
laboratory room temperature of 22~$^\circ$C: $T_{ion}=(31\pm9)$~$^\circ$C.

During operation, the \Al clock servo alternates between probing of the
\clocktransitionAlmF{5/2}{5/2} and \clocktransitionAlmF{-5/2}{-5/2}
transitions every 5~s, and the apparatus synthesizes an average of
these two frequencies to eliminate first-order Zeeman shifts~\cite{Madej1998, TR2007Al3P0observed}.
Each transition's resonance is probed several times at the high- and low-frequency half-maximum points to derive a frequency-correction signal.  The frequency-difference between the transitions is proportional to the
mean magnetic field $\langle B \rangle$, which allows an accurate
estimate of the quadratic Zeeman shift due to the quasi-static
quantization field of typically $\langle B \rangle = 0.1$ mT.  However,
the quadratic Zeeman shift is proportional to $\langle B^2\rangle=\langle B
\rangle^2+B_{AC}^2$, where $B_{AC}^2$ is the variance of the magnetic
field about its mean.  The dominant sources of varying magnetic fields
are currents at $\nu_{RF}$ in conductors near the ion.
We vary the trap RF drive power $P$ and measure the frequency of the hyperfine clock transition in \MgTF
$|F=3,m_F=0\rangle\rightarrow|F=2,m_F=0\rangle$ near 1.789~GHz, which has a
strong quadratic dependence on the magnetic field, to find
$B_{AC}^2=1.45\times10^{-12}(P/\textrm{W})$~T$^2$.  For \Al clock operation $P = 15$~W, and $B_{AC}^2 =
2.17\times10^{-11}\textrm{T}^2$, which alters the quadratic Zeeman shift by
$\Delta\nu/\nu=(-1.4\pm0.3)\times10^{-18}.$

Other potential systematic shifts are listed in Table~\ref{systematics}.
When stabilizing the clock laser to the ion we probe the clock transition
alternately with counter-propagating laser beams to observe and cancel
potential first-order Doppler shifts~\cite{TR2008AlHg}.
For the clock comparison described below, we observe a differential shift
for the two probe directions of $(1.2\pm0.7)\times10^{-17}$ but this effect is suppressed by taking the average.
The suppression factor is limited because the atomic line-shape and hence the servo gains differ slightly for the two probe directions.
During clock operation the mean fractional gain imbalance was 1.5~\%, thereby reducing the
possible first order Doppler shift to $3\times10^{-19}$.

We have looked for Stark shifts due to the clock
pulse itself by raising the intensity of the probe beam in one clock, and comparing the
frequency to that produced by the other \Al clock.  With an increase of 40 dB in the clock probe
intensity, we observe no statistically significant frequency difference
with a fractional uncertainty of $2\times10^{-15}$,  constraining any effect that scales linearly
with this intensity to $2\times10^{-19}$.
We observe a rate of background gas collisions similar to the Al-Be
clock and assign a $5\times10^{-19}$ uncertainty to this potential shift~\cite{TR2008AlHg}.
Thermally induced frequency errors from beam switching AOMs (Fig.~\ref{Setup})
have been evaluated previously~\cite{TR2008AlHg}, and this
uncertainty is reduced to $2\times10^{-19}$ by operating the AOMs at less than 1 mW.

\begin{figure}
\includegraphics[angle=0, width=0.5\textwidth]{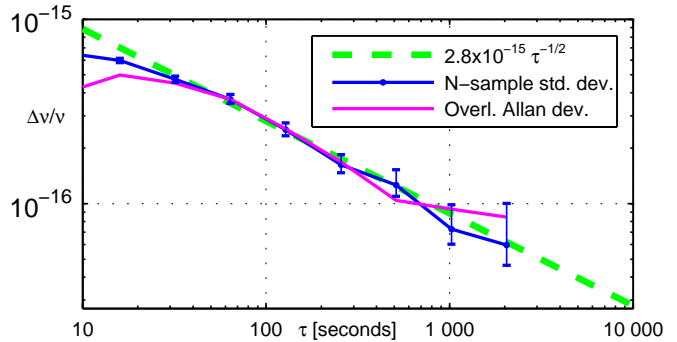}
\caption[Stability.]{Clock stability. Fractional frequency uncertainty vs. averaging period
($\tau$) for a comparison between the two Al$^+$ clocks (10700 s duration).
Overlapping Allan deviation and N-sample standard deviation are shown~\cite{Riley2008Stability}.
For each comparison measurement the coefficient of the $\tau^{-1/2}$
asymptote is estimated and used to derive the measurement's statistical uncertainty.
The $2.8\times10^{-15}\tau^{-1/2}$ asymptote is reached for averaging periods that are longer than the servo time constant of 10 s.}
\label{Stability}
\end{figure}

\begin{figure}
\includegraphics[angle=0, width=0.48\textwidth]{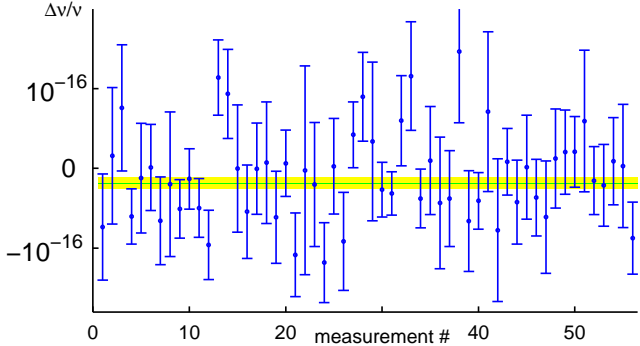}
\caption[Results.]{Measurements of the fractional frequency difference between the
two \Al clocks (blue points).  Error bars represent the statistical uncertainty (see Fig.~\ref{Stability}).  The horizontal line shows the weighted mean of \diffAlAl with an overall statistical uncertainty of $\pm$\uncStat (shaded band).}
\label{Results}
\end{figure}

We directly compared the Al-Mg clock with
the Al-Be clock to perform independent tests that could reveal unaccounted-for clock shifts.
The Al-Be clock was evaluated with an accuracy of
\systAlBe~\cite{TR2008AlHg}, and this evaluation remained valid
during the two-clock comparison.  The two clocks were compared
in 56 separate measurements, each of duration $1000-11000$~s.
As shown in Fig.~\ref{Setup}, the frequency of the \clocktransition
probe laser was actively steered to the \Al ion in the Al-Mg apparatus
with a servo time-constant of about 10 s.  A portion of this laser light
simultaneously probed the \Al ion in the Al-Be clock, where it was
servoed to the clock transition in a separate digital feedback loop.  The frequency
produced by this feedback loop represents the frequency difference
between the two clocks, and was recorded and analyzed for
stability (Fig.~\ref{Stability}).  Average frequencies of the
individual measurements corrected for known shifts are shown in
Fig.~\ref{Results}, where the overall weighted mean is
$(\nu_{AlMg}-\nu_{AlBe})/\nu=$ \diffpmAlAlns.  This value is consistent with the
1-$\sigma$ error of $2.5\times10^{-17}$ that is calculated by adding in quadrature the inaccuracies of the two clocks
and the statistical uncertainty.  The reduced-$\chi^2$
for this data set is $1.02$, indicating that the error bars derived from the estimated stability correctly capture the scatter of the data.  For this data set the total in-loop servo error for the Al-Mg clock was $\dnn=6\times10^{-19}$, and for a well-designed servo loop this error declines faster than the statistical error.  It is therefore not included as a systematic shift in Table~\ref{systematics}.

In summary, we have built an \Al ion clock with a fractional frequency inaccuracy of \systAlMgns.
Its frequency is compared
to that of a previously constructed \Al clock, and the measured fractional frequency
difference of \diffpmAlAl is consistent with the inaccuracy of the previous clock.
Significantly, the statistical uncertainty in the frequency comparison of \uncStat is smaller than the
inaccuracy of either clock, and the average measurement stability was $2.8\times10^{-15}\tau^{-1/2}$  (total measurement duration: 164~967 s).  This result may be compared to other direct same-species atomic clock comparisons, where two Cs microwave clocks have reached an agreement of $(4\pm3)\times10^{-16}$~\cite{Vian2005Cs}, and two Yb$^+$ single ion clocks showed agreement of $(3.8\pm6.1)\times10^{-16}$~\cite{Schneider2005}.  Future frequency ratio measurements of the \Al
clock and the NIST \Hg optical clock would enable improved constraints
on present-era changes in the fine-structure constant~\cite{TR2008AlHg}.

We thank J.~C.~Bergquist for use of his stable Fabry-Perot cavities, T.~M.~Fortier for optical frequency
measurements, W.~M.~Itano for atomic structure calculations, and S.~Bickman and D.~Leibrandt for helpful comments on the manuscript.
This work was supported by ONR. J.C.J.K. acknowledges support from the Netherlands Organisation for Scientific Research (NWO).
Contribution of NIST, not subject to U.S. copyright.


\begin{thebibliography}{10}
\bibitem{TwoLaserGSC}%
  \BibitemOpen
  \bibinfo {note} {The entropy-reducing laser that repumps the states
  $|^2S_{1/2},F\in\{2,3\},m=-2\rangle$ via $|^2P_{1/2},F=3,m=-3\rangle$ into
  $|^2S_{1/2},F=3,m=-3\rangle$ is replaced by the repeated application of a
  pulse sequence that first drives the
  $|^2S_{1/2},F=2,m=-2\rangle\rightarrow|^2P_{3/2},F=3,m=-3\rangle$ transition
  and subsequently the transition
  $|^2S_{1/2},F=3,m=-2\rangle\rightarrow|^2S_{1/2},F=2,m=-2\rangle$.}%
  \bibAnnoteFile{Stop}{TwoLaserGSC}%
\end{thebibliography}

%

\end{document}